\documentstyle[preprint,aps,pra]{revtex}

\def\eps{\epsilon}

\begin{document}
\title{Condensate  of  charged Bose disks
 with numerous holes in a uniform magnetic field}
\author{Sang-Hoon Kim}
\address{Division of Liberal Arts, Mokpo National Maritime University,
 Mokpo 530-729, Korea}
\date{\today}
\maketitle
\draft
\begin{abstract}
A stack of disks with numerous holes
 composed of a non-interacting charged Bose gas
is modeled as a low dimensional disk.
The Bose condensation of the net-like disk system
in a uniform magnetic field is studied.
Calculation of the condensate fraction of the net-like disk system  
showed that there still exists a non-zero condensate fraction
 at low temperature and in weak field.
\end{abstract} 
\newpage

It is known that two fermions can be coupled to form a
 pair, which  behaves like a spinless boson.
Many bosonic pairs form a kind of charged Bose gas, 
and it has also been
 known that the condensation of the Bose system  could be 
a reliable candidate for superfluidity\cite{alex1,alex4}.
It has also been reported that a dominant contribution to 
the superfluid 
density of liquid helium-4 in films and porous media originates from 
a geometric  structure, such as the non-integer dimensionality
 of the samples\cite{kim}. 
Condensate density plays a key role in the superfluid density.
In this note we will discuss a condensation of an odd 
structure of matter, a system of charged bose disks (CBD).

We assume that the distance  between any two fermions
 of a pair is large enough to neglect the Coulomb interaction.
Then, we propose a stack of
 non-interacting low dimensional charged Bose disks
with numerous holes in a uniform magnetic field.
In reality, it is a very thin net  
with negligible thickness and numerous holes. 
Fig. 1 is a simple example of the CBD. 
Therefore, the net could be modeled into a low dimensional
 disk which has a dimension between 1 and 2 
in fractal point of view\cite{pie,bun}.
The dimensionality represents the effects in the measure of
{\it disorderness} in terms of the connectivity and complexity
 of the system\cite{PfOb,Pf}.

Generally, magnetic field hardly penetrates superconductor, but
can pass through the net-like disk without any difficulty.
Therefore, the low dimensional CBD may not be antiferromagtic.
This is an advantage of the low dimensional CBD.
It  has been known that an ideal charged Bose gas in two 
dimensions (2D) cannot be condensed  under a magnetic field 
because of the one dimensional (1D) character of particle motion
within the lowest Landau level\cite{scha,alex3,raso}. 
On the other hand, if we pile up the net-like low
 dimensional disks in  parallel, this gives an extra dimension to
 the perpendicular axis.
The whole dimension of the stack then
 becomes between 2 and 3, which  is large  enough to
 create a nonzero superfluid density. 

We apply  a simple statistical approach for the $D$ non-integer
 dimensions.
The theory we use begins from the non-interacting Bose gas 
in disk dimensions. It is uniform in disk directions.
 This is then extended to a charged system such as the
bipolaronic method for the condensate  density\cite{alex1,alex4}.
The  partition function of the system is given by
\begin{equation}
\ln {\mathcal{Q}_D}(z,T) = -\int_{0}^{\infty} d\eps \, \rho_D(\eps)
\ln (1-z e^{-\eps /T}) - \ln (1-z),
\label{11}
\end{equation}
where  $z$ is the fugacity defined by $z=e^{-|\mu|/T}$, 
and $\eps_{\bf p} = \frac{p^2}{2m}$ is taken for the neutral system.
We set  $\hbar=c=k_B =1$ for convenience, and unit volume is assumed.

The term $\rho_D(\eps)$ is the $D$-dimensional density of states
and plays a key role in our analysis.  For a neutral and uniform system
 it is  given as \cite{kim,PfOb,Pf}
\begin{equation}
\rho_D(\eps) = a_D \eps^{\frac{D}{2} - 1},
\label{14}
\end{equation}
where $a_D$ is a $D$-dimensional coefficient given by  
$a_D =  \Gamma\left( \frac{D}{2} \right)^{-1}
\left( \frac{m^\ast}{2 \pi} \right)^{\frac{D}{2}}.$
Here, $\Gamma$ is the Gamma function and $m^\ast$ is
 the effective mass of a pair.

The average number of particles is obtained from Eq. (\ref{11})
\begin{eqnarray}
n &=& z\frac{\partial \ln {\mathcal{Q}}}{\partial z}
\nonumber \\
&=&\int_0^\infty d\eps \frac{\rho_D(\eps)}{z^{-1} (e^{\eps /T} - 1)}
+ n_0,
\label{15}\end{eqnarray}
 where $n_0 = n_{{\bf p}=0}$. 
Next, substituting the $D$-dimensional
density of states from Eq. (\ref{14}) into Eq. (\ref{15}),
the condensate fraction is obtained as\cite{kim,Pf,huan}
\begin{eqnarray}
\frac{n_0}{n} &=&
1- v \int_0^\infty d\eps \frac{\rho_D(\eps)}{z^{-1}
 (e^{\eps /T} - 1)}
\nonumber \\
&=& 1- \int_0^\infty d\eps \frac{\eps^{\frac{D}{2}-1}}{e^{\eps /T} - 1}
\left[ \int_o^\infty d\eps \frac{\eps^{\frac{D}{2}-1}}{e^{\eps /T} - 1}
\right]_{T_c}^{-1}
\nonumber \\
&=& 1-  \left\{ \frac{T}{T_c(D)} \right\}^{\frac{D}{2}},
\label{16}
\end{eqnarray}
where
\begin{equation}
T_{c}(D) = \frac{2 \pi }{m^\ast v^{\frac{2}{3}}}
\frac {1}{\zeta(\frac{D}{2})^{\frac{2}{D}}}.
\label{ctem}
\end{equation}
Here, $v$ is the volume density and $\zeta$ is the Riemann-Zeta function.
The $z=1$ limit is taken for the condensation.
Note that $\int_0^\infty dx \, x^{\frac{D}{2} -1}/(e^x - 1) 
= \Gamma(\frac{D}{2}) \zeta(\frac{D}{2}).$

The critical temperature in Eq. (\ref{ctem}), $T_c$,  
 corresponds to the BEC transition temperature.
It is rewritten as a function of  $T_c^b$  for the bulk $(D=3)$
 as
\begin{equation}
T_c(D) = \alpha(D)  T_c^b,
\label{40}
\end{equation}
where $\alpha(D) = 1.897/ \zeta \left( \frac{D}{2}\right)^{2/D}$.
Note that $\zeta(\frac{3}{2})^{\frac{2}{3}} = 1.897$.
 It can be readily shown that 
 Eq. (\ref{40}) satisfies both the ideal thin limit ($D=2$)
 and the bulk limit ($D=3$).
Note also that the transition would not occur for the 2D limit since
 $ T_c  \sim  \left| \frac{D}{2}-1 \right|$
 as  $D$ approaches to 2\cite{Grad}.

The CBD in a uniform magnetic field is now extended
by using this new $D$-dimensional density of states, $\rho_D(\eps,\omega)$.
Our $D$-dimensional system, $2<D<3$, is composed of $(D-1)$-dimensional
net-like planes and an additional dimension that is parallel to 
the magnetic field.
A uniform magnetic field is applied  perpendicular to the direction
of the disks. 
The new density of states in $D$-dimensions is derived 
from the Landau quantization law\cite{alex1,land}:
$ \eps_{n}+\eps_{p_z}  
= \left( n+\frac{1}{2} \right)\omega_H + p_z^2 /2 m^\ast $,
where $\omega_H=2eH/m^\ast$ is the cyclotron frequency.

	The $(D-1)$-dimensional degeneracy is
\begin{equation}
\rho_{D-1}(\eps) \omega_H = \frac{1}{\Gamma\left(\frac{D-1}{2}\right)}
\left( \frac{m^\ast}{2 \pi}\right)^{\frac{D-1}{2}}
\eps^{ \frac{D-3}{2} }\omega_H,
\label{43}
\end{equation}
Applying the Landau quantization energy 
to the density of states of the CBD,
we obtain  $\rho_D(\eps,\omega_H)$  as
\begin{eqnarray}
\rho_D(\eps,\omega_H) &=& \rho_{D-1}(\eps) \omega_H
\sum_{n=0}^{\infty}\sum_{p_z} \delta(\eps - \eps_{n} - \eps_{p_z}).
 \nonumber \\
 &=& \rho_{D-1}(\eps) \omega_H
 \sum_{n=0}^{\infty}\int_{-\infty}^{\infty}
  \frac{d p_z}{2\pi} \delta(\eps - \eps_{n}-\eps_{p_z})
 \nonumber \\
&=& \frac{1}{\sqrt{\pi} \, \Gamma\left(\frac{D-1}{2}\right) }
\left( \frac{m^\ast}{2 \pi}\right)^{\frac{D}{2}}
\eps^{\frac{D-3}{2}}\omega_H
\sum_{n=0}^\infty \frac{1}{ \sqrt{\eps - (n+\frac{1}{2})\omega_H}}.
\label{45}
\end{eqnarray}
Certainly, this $\rho_D$ of the net-like system is different from
the one of uniform system in Eq. (\ref{14}) 

Therefore, the condensate fraction of the net-like system is
 obtained as
\begin{eqnarray}
\frac{n_0}{n}(T,\omega_H)
&=& 1- 
 \int_o^\infty d\eps \frac{\rho_D(\eps,\omega_H)}{e^{ \eps /T} - 1}
\left[ \int_o^\infty d\eps \frac{\rho_D(\eps,\omega_H)}{e^{ \eps /T} - 1}
 \right]_{T_c}^{-1}
\nonumber \\
&=&
1- \frac{\int_0^\infty  d\eps \frac{\eps^{\frac{D-3}{2}}}{e^{\eps/T}-1}
\sum_{n=0}^\infty 
\frac{1}{\sqrt{\eps - (n+\frac{1}{2}) \omega_H}}}
{\int_0^\infty  d\eps  \frac{\eps^{\frac{D-3}{2}}}{e^{\eps/T_c(D)}-1}
\sum_{n=0}^\infty 
\frac{1}{\sqrt{ \eps - (n+\frac{1}{2}) \omega_H}} }.
\label{46}
\end{eqnarray}

We may use Eq. (\ref{40}) for $T_c$, and 
introduce the dimensionless variables $t$ and $y$ instead
of $T$ and $\omega_H$ as
$t=\frac{T}{T_c^b}$ and $y = \frac{\omega_H}{T_c^b}$.
 Then, the condensate fraction is expressed in the following
simple form 
\begin{equation}
\frac{n_0}{n}(t,y)
= 1-  \left(\frac{t}{\alpha}\right)^{\frac{D}{2}-1}
\frac{A(y/t)} {A(y/\alpha)}.
\label{47}
\end{equation}
 $A(y)$  is defined by 
\begin{equation}
A(y)= \int_{x_0}^\infty dx \frac{x^{\frac{D-3}{2}}}{e^x - 1}
\sum_{n=0}^\infty
\frac{1}{\sqrt{x-(n+\frac{1}{2}) y}},
\label{48}
\end{equation}
where $x_0= (n+\frac{1}{2}) y $. 
$A(y)$ itself can diverge, but $n_0/n$ in 
Eq. (\ref{47}) does not.

The condensate density of the charged Boson model should be different
from Eq. (\ref{16}) as $D\rightarrow 3$ limit, and contain
 additional factor of $A(y)$ which gives the effect of the 
field over that of the neutral model in Eq. (\ref{16}).
From the equation above, we can easily check that the condensate
 fraction goes to zero at $t>\alpha$  and $y>1$ region.
On the other hand, we also know that there still exists 
 non-zero condensate fraction at low temperature
 and low frequency range, but it quickly drops to zero 
as the field or temperature is increase.

It is plotted in FIG. 2, at various disk dimension and frequencies
as a function of normalized temperature $t$.
The graphs in Fig. 2 are irregular since the summation of $n$
is  contained in the integral,
 but the over all shape of the drop is clear.
We see that the higher dimension gives a little bit more 
condensate fraction.

We calculated the density of states and
condensate fraction of the net-like CBD in the uniform
 magnetic field, and find that the field strongly effects the
 condensate fraction of the CBD.
The condensate fraction drops to zero quickly as the temperature
and field increase, 
 but it still exists at the temperature of $t <\alpha$
and in  the  field of $y<1$.
We think that the numerous holes of the CBD help
the condensate fraction to keep non-zero at the range.

We send special thank to  A. S. Alexandrov,  C. K. Kim, and 
P. Pfeifer for fruitful advice.

\begin{figure}
\caption{ A net-like model of charged Bose disks with numerous holes.}
\end{figure}
\begin{figure}
\caption{The condensate fraction  of the charged Bose disks 
in a uniform magnetic field.
The solid line is for $y=10$, the dashed line is for $y=4,$
and the dotted line is for $y=1.$
(a) When D=2.4 (the dimension of a disk is 1.4). 
(b) When D=2.8 (the dimension of a disk is 1.8).}
\end{figure}
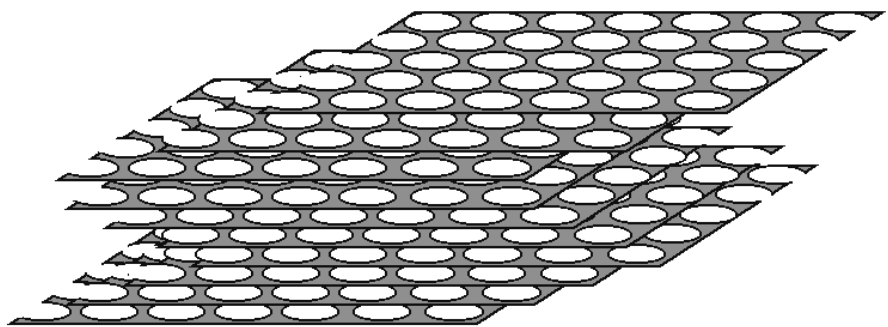
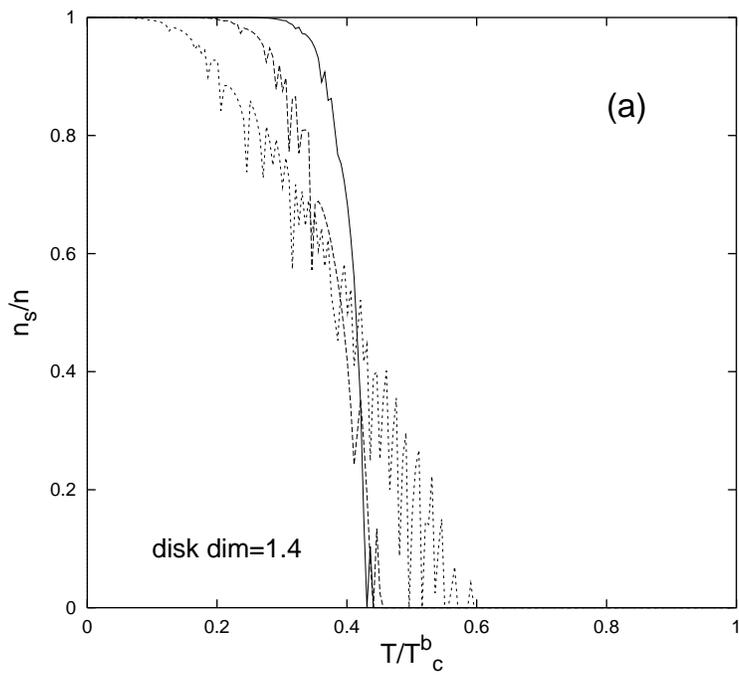
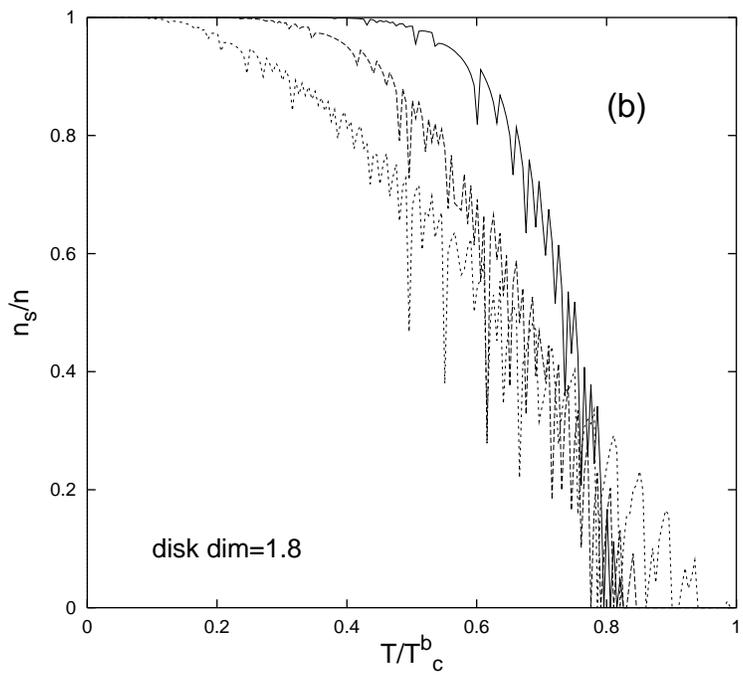
\end{document}